\documentclass[aps,pre,twocolumn,showpacs,a4paper,superscriptaddress]{revtex4-1}
\usepackage{graphicx,bm,amsmath,dcolumn,amssymb}
\usepackage{longtable}
\graphicspath{{figures}}
\usepackage{geometry,hyperref,color}
\begin{document}

\title{Monte Carlo simulation of quantum Potts model}
  \author{Chengxiang Ding}
  \email{dingcx@ahut.edu.cn}
  \affiliation{School of Science and Engineering of Mathematics and Physics, Anhui University of Technology, Maanshan 243002, China }
  \author{Yancheng Wang}
  \affiliation{Beijing National Laboratory for Condensed Matter Physics, and Institute of Physics, Chinese  Academy of Sciences, Beijing 100190, China}
  \author{Youjin Deng}
  \affiliation{Hefei National Laboratory for Physical Sciences at Microscale,
  Department of Modern Physics, University of Science and Technology of China, Hefei, 230027, China}
  \author{Hui Shao}
  \affiliation{Beijing Computational Science Research Center, Beijing 100193, China}
 \affiliation{Department of Physics, Boston University, 590 Commonwealth Avenue, Boston, Massachusetts 02215, USA}
\date{\today}
\begin{abstract}
Using Monte Carlo simulations in the frame of stochastic series expansion (SSE), we study the three-state quantum Potts model.
The cluster algorithm we used is a direct generalization of that for the quantum Ising model.
The simulations include the one dimensional and two dimensional ferromagnetic three-state quantum Potts model and the two dimensional antiferromagnetic
three-state quantum Potts model.
Our results show that the phase transition of the one dimensional ferromagnetic quantum Potts model belongs to 
the same universality class of the two dimensional classical Potts model,
the two dimensional ferromagnetic quantum Potts model undergoes a first order transition, which is also in analogy to its classical correspondence.
the phase transition of the antiferromagnetic quantum Potts model is continuous, 
whose universality class belongs to the three-dimensional classical $XY$ model, owing to an `emergent' O(2) symmetry at the critical point, 
although its ordered phase breaks the $Z_6$ symmetry.
\end{abstract}
\pacs{05.50.+q, 64.60.Cn, 64.60.Fr, 75.10.Hk}
\maketitle 
\section{Introduction}
The Potts model\cite{Potts}, as a generalization of the famous Ising model, plays an important role in the study of phase transition and critical phenomena, 
which shows a typical phase transition of symmetry breaking. This model is widely studied, including the ferromagnetic case and antiferromagnetic case.
However, the research of the quantum version of the model, i.e., the quantum Potts model\cite{quanPotts} is relatively few\cite{qp2014,qp2015}. This seems somewhat strange, 
comparing to the fact that the quantum version of the Ising model, namely the quantum Ising model or the transverse field Ising model, 
has been studied extensively\cite{kagomeTIM,FFTIM,FTTIM,triangleTIM,squareice,trilayerIsing}.

The quantum phase transition of the one dimensional ferromagnetic quantum Potts model is well understood\cite{Baxter1973,quanPotts}, 
the critical point $g_c=1$ for all $q$, 
and the universality class being same as the corresponding two dimensional classical Potts model, which is first order for $q>4$ and continuous for $q\le 4$.
The two dimensional ferromagnetic quantum Potts model is studied by the infinite projected entangled-pair state (iPEPS) algorithm\cite{qp2014}, showing that 
both the 3-state and 4-state ferromagnetic quantum Potts models undergo a first-order phase transition. This result is easy to understand, because their classical 
correspondence, namely the classical three dimensional 3-state and 4-state ferromagnetic Potts models, also undergo first-order phase transitions.
For the antiferromagnetic quantum Potts model, to the best of our knowledge, the study is almost blank 
except the recent work on the mixed classical Potts model\cite{mixedPotts},
which can be considered as the classical correspondence of it.

In this paper, we will study the quantum Potts model by the means of Monte Carlo simulations in the frame of stochastic series expansion (SSE), using a cluster algorithm
which is a direct generalization of the cluster algorithm for quantum Ising model. Our study include both the ferromagnetic case and antiferromagnetic case. 
\section{model and method}
The Hamiltonian of the quantum Potts model\cite{quanPotts} can be written as
\begin{eqnarray}
H=-J\sum\limits_{<i,j>}\sum\limits_{k=1}^{q-1}\Omega_i^k\Omega_{j}^{q-k}-g\sum\limits_i\sum\limits_{k=1}^{q-1}\Gamma_i^k
\end{eqnarray}
where $\Omega$ and $\Gamma$ are $q\times q$ matrices. $\Omega$ is a diagonal one, 
\begin{equation}      
\Omega=\left(           
  \begin{array}{ccccccc}   
    1 & & & & &  \\
      &\omega & & & &  \\
      & &\omega^2& & &  \\
      & & &\cdot & &  \\
      & & & &\cdot &  \\
      & & & & &\cdot  \\
      & & & & & &\omega^{q-1}  \\
  \end{array}
\right),
\quad \omega=e^{\frac{2\pi i}{q}},
\end{equation}
and $\Gamma$ is an off-diagonal one, 
\begin{equation}
\Gamma=\left(           
  \begin{array}{cc}   
   0 & I_{q-1}\\
   1 & 0
  \end{array}
\right),
\end{equation}
where $I_{q-1}$ is a $(q-1)\times(q-1)$ identity matrix.
This model is ferromagnetic for $J=1$ or antiferromagnetic for $J=-1$. $g\ge0$ is the strength of the 
transverse field.  When $q=2$ it reduces to the quantum Ising model. 

The exponential of the partition function $Z=\rm{Tr}\{\exp{(-\beta H)}\}$
in SSE method\cite{SSE,SSE1} is written as a Taylor expansion 
\begin{eqnarray}
Z=\sum\limits_{\alpha_0}\big\langle \alpha_0\big|\sum\limits_{n=0}^\infty\frac{\beta^n}{n!}(-H)^n\big|\alpha_0\big\rangle
\end{eqnarray}
where $|\alpha_0\rangle=|\sigma_1,\sigma_2,\cdots,\sigma_N\rangle$ is  the standard spin basis.
Then, by inserting a set of complete basis, the partition function is rewritten as 
\begin{eqnarray}
Z=\sum\limits_{\{\alpha_p\}}\sum\limits_{n=0}^\infty\frac{\beta^n}{n!}\prod\limits_{p=0}^{n-1}\langle\alpha_p|(-H)|\alpha_{p+1}\rangle
\end{eqnarray}
with $\alpha_n=\alpha_0$.

\begin{figure}[htpb]
\includegraphics[scale=0.6]{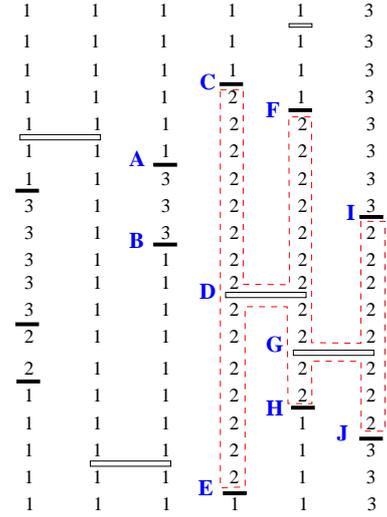}
\caption{A typical SSE configuration of the three-state quantum Potts model.}
\label{ssecfg}
\end{figure}

In the next step, the Hamiltonian is written as a sum of elementary lattice operators
\begin{eqnarray}
H=-\sum\limits_t\sum\limits_a H_{t,a}
\end{eqnarray}
the subscript $t$ and $a$ refer to the type and the position of the operators respectively.
Therefore, the partition function can be written as 
\begin{eqnarray}
Z=\sum\limits_{\{\alpha_p\}}\sum\limits_{n=0}^\infty\sum\limits_{S_n}\frac{\beta^n}{n!}\prod\limits_{p=0}^{n-1}\langle\alpha_{p}|H_{t_p,a_p}|\alpha_{p+1}\rangle \label{sse1}
\end{eqnarray}
with $S_n=[t_1,a_1]$, $[t_2,a_2]$,$\cdots$, $[t_n,a_n]$ the sequence of operators.
The Taylor expansion in (\ref{sse1}) can be truncated at certain power $M$ for facilitating the implementation of the updating scheme. If $M$ is large enough, 
the truncation error should be completely negligible.
After the truncation, the partition function is written as
\begin{eqnarray}
Z=\sum\limits_{\{\alpha_p\}}\sum\limits_{S_M}\frac{\beta^n(M-n)!}{M!}\prod\limits_{p=0}^{M-1}\langle\alpha_p|H_{t_p,a_p}|\alpha_{p+1}\rangle\label{sse}
\end{eqnarray}
The binomial coefficient $M!/(M-n)!n!$ comes from the random insertion of the $M-n$ unit operators(which is often called `vacant' operator in SSE simulations), 
because $n\le M$. (\ref{sse}) is the starting point of the SSE Monte Carlo simulation, it defines the `SSE configuration' which consists of a sequence of states 
$|\alpha_p\rangle$=$|\sigma_1(p)$, $\sigma_2(p)$, $\cdots$, $\sigma_N(p)\rangle$ with $p=0$ to $M$ (with $|\alpha_0\rangle$=$|\alpha_M\rangle$) and operators,
whose weight is 
\begin{eqnarray}
 \frac{\beta^n(M-n)!}{M!}\prod\limits_{p=0}^{M-1}\langle\alpha_p|H_{t_p,a_p}|\alpha_{p+1}\rangle,\label{ssewgt}
\end{eqnarray}
Fig. \ref{ssecfg} shows a typical SSE configuration.
It should be noted that it is not necessary to store all the states in the propagation direction, the information of $|\alpha_0\rangle$ and the sequence of operators 
can generate all the states.

In the quantum Potts model, the local basis (spin states) can be chosen as the eigenstates of the diagonal matrix $\Omega$, take the case of $q=3$ for example, the local spin
states are 
\begin{eqnarray}      
|1\rangle=\left(           
  \begin{array}{c}   
    1\\
    0\\
    0
  \end{array}
\right),
|2\rangle=\left(           
  \begin{array}{c}   
    0\\
    1\\
    0
  \end{array}
\right),
|3\rangle=\left(           
  \begin{array}{c}   
    0\\
    0\\
    1
  \end{array}
\right),
\end{eqnarray}
\begin{figure}[htpb]
\includegraphics[scale=0.4]{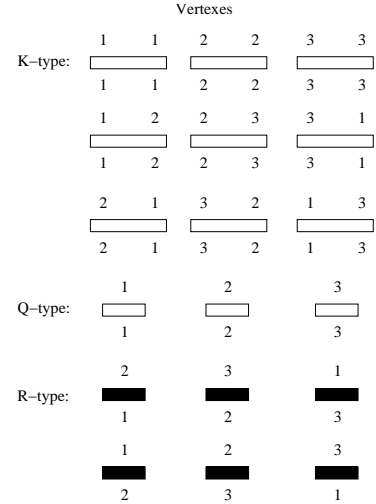}
\caption{SSE vertexes of the three-state quantum Potts model. the $K$-type and $Q$-type vertexes are diagonal, the $R$-type vertex is off-diagonal.}
\label{ssevertex}
\end{figure}
and the local Hamiltonian $H_{t,a}$ includes the diagonal operator $K_{i,j}=J(\Omega_i\Omega_j^2+\Omega_i^2\Omega_j)$ 
 and the off-diagonal operator (transverse field) $R_i=g(\Gamma_i+\Gamma_i^2)$. 
 For facilitating the simulation, we further define a diagonal operator $Q_i=g$. 
 This operator also shifts the energy of the system but does not change the substantial physical properties of the model.
 The `vertex' for the SSE simulation can be defined based on the operator and the states of the spins connected to it,
 whose weight is $\langle\alpha_p|H_{t_p,a_p}|\alpha_{p+1}|\rangle$. 
 Corresponding to the operators, there are three types of vertexes, as shown in Fig. \ref{ssevertex}.
 The weight of the $K$-type vertex is 
\begin{eqnarray}
\langle \sigma_i,\sigma_j|K_{i,j}|\sigma_i,\sigma_j\rangle=J(-1+q\delta_{\sigma_i,\sigma_j}).\label{Kwt}
\end{eqnarray}
When $\sigma_i\ne\sigma_j$, the weight is negative, which will lead to sign problem, thus in the simulations
this weight is set as $C+J(-1+q\delta_{\sigma_i,\sigma_j})$, which only shifts the energy of the system, but avoids the sign problem if the value of $C$ is set properly.
In the cluster algorithm that we will describe below, this value must be set as $J$ for the ferromagnetic case ($J$=1) and $-J(q-1)$ for the antiferromagnetic case ($J$=-1).
The weights of the $R$-type and $Q$-type vertexes are
\begin{eqnarray}
\langle \sigma(p)|R| \sigma(p+1)\rangle&=&g(1-\delta_{\sigma(p),\sigma(p+1)}),\\
\langle \sigma(p)|Q| \sigma(p+1)\rangle&=&g\delta_{\sigma(p),\sigma(p+1)}.\label{Qwt}
\end{eqnarray}
The weights of the two types of vertexes are complementary to each other, which allows a free
change of the two types of vertexes to each other in the procedure of simulation. This is very important for the realization of the off-diagonal update 
of the SSE configuration, including both the Metropolis algorithm and the cluster algorithm.

The simulation of Eq. (\ref{sse}) consists of the `diagonal update' and `off-diagonal update'. 
For the diagonal update, the procedure consists of inserting a diagonal vertex to the sequence or removing a existing one.
Both the $K$-type and $Q$ type vertexes are diagonal, and the Metropolis procedure is performed as
\begin{enumerate}
\item if the vertex is vacant, the probability of trying to insert a $K$-type vertex is $\xi$, accordingly, the probability of trying to insert a $Q$-type vertex
is $1-\xi$. The acceptance probability of inserting the $K$-type vertex is
\begin{eqnarray}
P=\frac{N_b\beta\langle\alpha_p|H_{t_p,a_p}|\alpha_{p+1}\rangle}{(M-n)\xi}. \label{diag1}
\end{eqnarray}
Here $N_b$ is the total number of bonds of the lattice. The acceptance probability of inserting a $Q$-type vertex is 
\begin{eqnarray}
P=\frac{N\beta g}{(M-n)(1-\xi)}. \label{diag2}
\end{eqnarray}
\item if the vertex is a diagonal one, then if it is $K$-type, the acceptance probability of removing it is 
\begin{eqnarray}
P=\frac{(M-n+1)\xi}{N_b\beta\langle\alpha_p|H_{t_p,a_p}|\alpha_{p+1}\rangle},
\end{eqnarray}
otherwise (it must be $Q$-type), the acceptance probability of removing it is 
\begin{eqnarray}
P=\frac{(M-n+1)(1-\xi)}{N\beta g}.
\end{eqnarray}
\end{enumerate}
The selecting probability $\xi$ can be adjusted to make the acceptance probability of inserting or removing to be a moderate one.

The off-diagonal update does not change the number of vertexes, but changes the type of vertexes.
In hard-core boson or Heisenberg model or some other models, a loop algorithm is used, which is very efficient. For the quantum Potts model,
two algorithms can be used for this purpose. The first one is a local updating algorithm (Metropolis),
which is based on the fact that the change of the $R$-type vertex to $Q$-type vertex
or conversely does not change the weight of the SSE configuration. For example, in Fig. \ref{ssecfg} change of the state of the series of spins
between the vertexes A and B (the state is $|3\rangle$) does not change the weight of the configuration;
the state can be randomly given to be one of the $q$ Potts states and the type of vertexes $A$ and $B$ should be changed
from $R$-type to $Q$-type, if the state is changed to $|1\rangle$.
Such algorithm is very easy to be implemented in compute programming.

However, the efficiency of the Metropolis algorithm is not satisfactory, especially for the large system at or near the critical point.
which suffers from the problem of critical slowing down. Therefore the simulations call for a nonlocal algorithm. 
Such algorithm is available for our model. In Ref. \cite{SSE-Ising}, a cluster algorithm is proposed for the quantum Ising model, which can be directly 
generalized to the quantum Potts model. In such algorithm, the construction of a cluster is started from a randomly selected leg of one vertex (not vacant), for example, 
a leg of a $K$-type vertex, then all the four legs and their links are considered to be in the same cluster and stored in the stack,
then the links of the legs are used to be the new stating point (four branches) of the cluster constructing procedure.
If the link is one of the leg of a $R$-type or $Q$-type operator, the constructing procedure terminates in such branch. 
The constructing procedure continues until all the possible branches terminates, then all the sates of spins of the cluster is flipped at the same time.
For the quantum Potts model, the construction of the cluster is {\it exactly} the same as the quantum Ising model, 
but the spin state of the cluster is randomly set as one of the $q$ Potts states; we also call this procedure as `flip'.
As mentioned earlier, the const $C$ is set as $J$ for the ferromagnetic model, therefore the weights of the vertexes as shown in the second and third 
rows of Fig. \ref{ssevertex} are zero, and the spins in the same cluster must have the same states.
In the numerical procedure, one can construct all the clusters and flip every cluster with probability 1/2 or construct {\it one} cluster and flip it with probability 1.
The former is similar to the Swendsen-Wang algorithm\cite{SW} and the later is similar to the Wolff algorithm\cite{Wolff}.

For the antiferromagnetic quantum Potts model, the diagonal update and the local algorithm for the off-diagonal update is the same as the ferromagnetic quantum Potts model,
but the cluster algorithm have to be modified. Take the three-state Potts model for example, in the cluster constructing procedure, we can let the spins with 
state $|1\rangle$ be frozen, and the cluster is constructed for the spins with value $|2\rangle$ or $|3\rangle$.
As mentioned earlier, the const $C$ is set as $-J(q-1)$, therefore the weights of the vertexes as shown in the first row of Fig. \ref{ssevertex} are are zero;
every pair of spins at the same side of a $K$-type vertex must have different states. After the construction, the state of the spin in the cluster should be changed 
from $|2\rangle$ to $|3\rangle$ or conversely (flip). Sincerely, we can also froze the spins with state $|2\rangle$, and flip the spins with state $|1\rangle$ or $|3\rangle$;
or froze $|3\rangle$, and flip $|1\rangle$ and $|2\rangle$. 
The ideal of such algorithm is very similar to the Wang-Swendsen-Koteck\'{y} algorithm\cite{3sAFP} of the classical antiferromagnetic Potts model.

\section{result}
\subsection{ferromagnetic case}
The temperature is set as $T=1/L$, namely $\beta=L$.
For the 3-state quantum Potts model, the simulated variables include the total energy $E$, the interaction  energy $E_z$, the  the magnetization $m$, 
and the Binder ratio $Q_m$
\begin{eqnarray}
E&=&-\frac{\langle n\rangle}{N\beta}+2C+g \label{En}\\
E_z&=&\bigg\langle\frac{1}{M}\sum\limits_{p=0}^{M-1} E_z(p)\bigg\rangle\\
m&=&\langle |\mathcal{M}|\rangle=\bigg\langle\bigg|\frac{1}{M}\sum\limits_{p=0}^{M-1}\mathcal{M}(p)\bigg|\bigg\rangle\label{mu}\\
Q_m&=&\frac{\langle\mathcal{M}^2\rangle^2}{\langle \mathcal{M}^4\rangle}\label{Qm}
\end{eqnarray}
with 
\begin{eqnarray}
 E_z(p)&=&\frac{-J}{N}\sum\limits_{<i,j>}(-1+q\delta_{\sigma_i(p),\sigma_j(p)})\\
\mathcal{M}(p)&=&\frac{1}{N}\sum\limits_{i=1}^N\vec{\sigma}_i(p).
\end{eqnarray}
Here, the states of the spins has been mapped to two dimensional vectors by the rule $\vec{\sigma}=(\cos\theta,\sin\theta)$ with $\theta=2\sigma\pi/3$.
In (\ref{En}), the const $2C+g$ is added to remove the energy shift that comes from the $K$-type and $Q$-type vertexes.

\begin{figure}[thpb]
\includegraphics[scale=0.7]{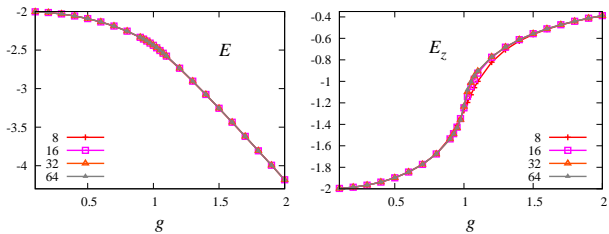}
\caption{Energy of the one dimensional three-state ferromagnetic quantum Potts model. Left, the total energy $E$; right, the interaction energy $E_z$.}
\label{E}
\end{figure}
\begin{figure}[htpb]
\includegraphics[scale=0.7]{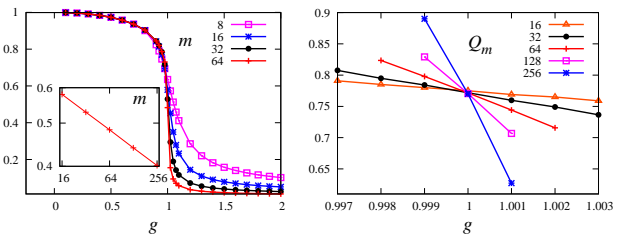}
\caption{The magnetization $m$ and Binder ratio $Q_m$ of the one dimensional three-state ferromagnetic quantum Potts model. 
The subset is the log-log plot of the magnetization $m$ versus system size $L$ with $g=g_c$.}
\label{m}
\end{figure}

Firstly, we simulate the one dimensional three-state quantum Potts model to test the algorithm. The energy $E$ and $E_z$ are in shown in the Fig. \ref{E}.
$E$ is featureless while $E_z$ obviously shows a phase transition. Such transition can also be detected by the magnetization $m$, as shown in the left of Fig. \ref{m}.
In order to determine the critical point and critical exponents of the phase transition,
we simulate the model near the critical point and fit the data of $Q_m$ (as shown in the right of Fig. \ref{m}) with
the finite-size scaling formula
\begin{eqnarray}
Q_m=Q_c+\sum\limits_{k=1}^2a_k(g-g_c)^kL^{ky_t}+bL^{y_u}, \label{Qmfss}
\end{eqnarray}
where $g_c$ is the critical point, $y_t>0$ the thermal exponent in the sense of renormalization, which is related to the critical exponent $\nu$ with $y_t=1/\nu$;
$\nu$ describes the divergence of the correlation length $\xi\sim|g-g_c|^{-\nu}$.
 $y_u$ is the exponent of the irrelevant field in renormalization, which is negative; $a_k$ and $b$ are unknown parameters; $Q_{c}$ is a shape-dependent universal parameter.
The fitting gives $Q_{c}=0.77(1)$, $g_c=0.99999(1)$, and $y_t=1.20(1)$. 
The estimated critical point and critical exponent coincides with the exact values\cite{quanPotts,Potts} $g_c=1$ and $y_t=6/5$. Exactly at the critical point, 
the value of $m$ scales with the system size with a power law formula
\begin{eqnarray}
m=L^{y_h-d}(a+bL^{y_u})\label{mfss}
\end{eqnarray}
as shown in the subset of Fig. \ref{m}.
Here, $y_h>0$ is the magnetic exponent that describes the renormalization of the magnetic field; it is related to the exponent $\eta$ with $y_h=2-\eta/2$;
 $\eta$ describes the critical behavior of the correlation function $G(r)\sim r^{-\eta}$.
The fitting gives $y_h=1.876(5)$, which coincides with the exact one $y_h=15/8$.
\begin{figure}[htpb]
\includegraphics[scale=0.7]{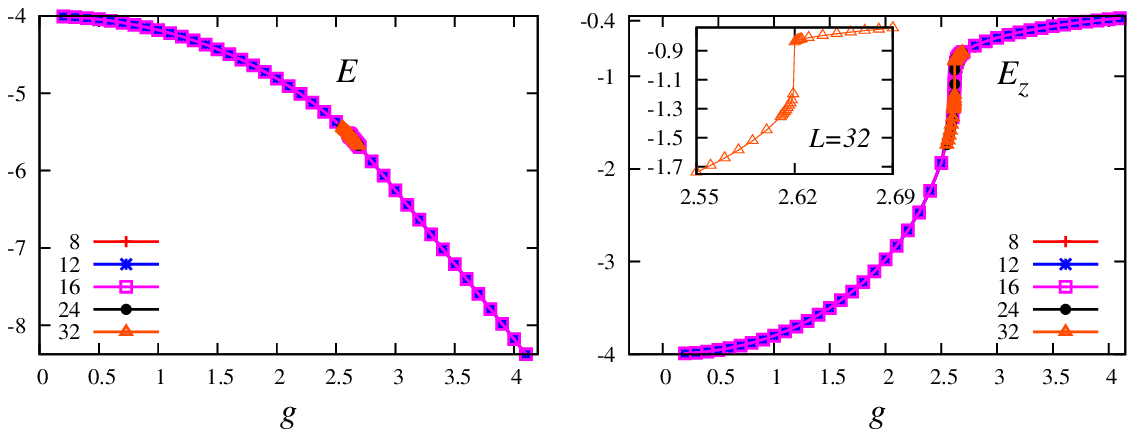}
\caption{Energy of the two dimensional three-state ferromagnetic quantum Potts model. Left, the total energy $E$; right, the interaction energy $E_z$.}
\label{E3sq}
\end{figure}
\begin{figure}[htpb]
\includegraphics[scale=0.75]{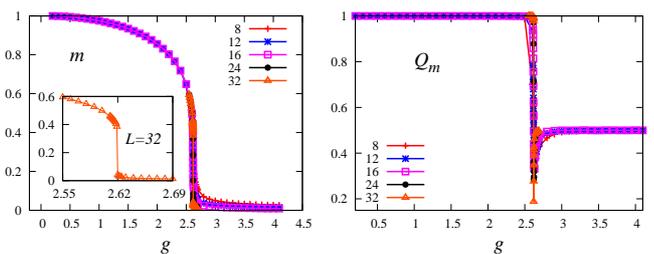}
\caption{Magnetization $m$ and the Binder ratio $Q_m$ of the two dimensional three-state ferromagnetic quantum Potts model.}
\label{m3sq}
\end{figure}

Now we turn to the two dimensional model.
The left of Fig. \ref{E3sq} shows the total energy of the three-state quantum Potts model on the square lattice, which does not show any singularity.
However the interaction energy $E_z$ shown in the right of Fig. \ref{E3sq} obviously show a gap, which is the typical feature of a first-order transition. 
Figure \ref{m3sq} shows the curves of $m$ and $Q_m$ versus $g$ for a series of $L$, it also obviously shows the phase transition;
the sharp falling of the magnetization the downward peak of the Binder ratio verifies that the transition is a first order one. 
From the behaviors of $E_z$, $m$, and $Q_m$, the transition point is estimated to be $g_c=2.619(3)$. 

\subsection{antiferromagnetic case}
We also simulate the antiferromagnetic quantum Potts model on the square lattice. In this case another order parameter called staggered magnetization is defined
\begin{eqnarray}
 m_s&=&\langle|\mathcal{M}_s|\rangle
\end{eqnarray}
with $\mathcal{M}_s=\mathcal{M}_A-\mathcal{M}_{B}$; 
$\mathcal{M}_A$ and $\mathcal{M}_B$ are defined by the similar way of (\ref{mu}),
but confined to the sublattices $A$ and $B$, respectively;
a lattice site belongs to sublattice $A$ or $B$ according to the parity of $x+y$, 
with $(x,y)$ the coordinate of the site. The Binder ratio of $m_s$ is defined by the similar way of (\ref{Qm})
\begin{eqnarray}
 Q_s=\frac{\langle \mathcal{M}_s^2\rangle^2}{\langle\mathcal{M}_s^4\rangle}
\end{eqnarray}

The transition of the antiferromagnetic quantum Potts model is found to be continuous.
The curves of the energy, the magnetization and the Binder ratio are similar to those of the one dimensional ferromagnetic quantum Potts model.
The data of $Q_s$ near the critical point can be fit by the formula (\ref{Qmfss}),
which gives $Q_c=0.72(1)$, $g_c=1.7173(3)$, and $y_t=1.48(1)$. The value of $y_t$ falls in the universality class of the three-dimensional classical $XY$ model;
to confirm this conclusion, we further simulate the model exactly at the critical point, and the data of $m_s$ can also be fit with the ansatz (\ref{mfss}).
The fitting gives $y_h=2.48(1)$, which is also consistent with the value of the three-dimensional classical $XY$ model.

\begin{figure}[htpb]
\includegraphics[scale=0.6]{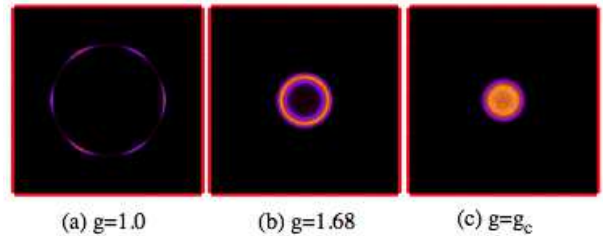}
\caption{Histograms of the staggered magnetization of the three-state antiferromagnetic quantum Potts model. The system size is $L=24$.}
\label{histo}
\end{figure}

However, it should be noted that the symmetry of the ordered phase of the model (with small $g$) is $Z_6$, which can be clearly demonstrated by the histogram of 
the staggered magnetization $(\mathcal{M}_{sx},\mathcal{M}_{sy})$, as shown in Fig. \ref{histo}(a).
The $Z_6$ symmetry is not difficult to understand. In the ordered phase, an ideal ordered configuration may be: the spins on sublattice A take states $|1\rangle$, while 
the spins on the sublattice B randomly take states $|2\rangle$ or $|3\rangle$. There are six types of the such ideal states, owing to the permutation
of the three states and the permutation of the two sublattices, therefore the symmetry of the order parameter is $Z_6$. 
The $XY$ universality class of the phase transition is owing to the `emergent' O(2) symmetry. Figure. \ref{histo}(b) and (c) give the histograms with
$g$ slightly below and exactly at the critical point respectively, which obviously demonstrates the O(2) symmetry.

The above results for the three-state quantum antiferromagnetic Potts model is very similar to the classical mixed Potts model\cite{mixedPotts} on simple cubic lattice, 
which can be considered as the classical correspondence of this model in the sense of universality.

\section{conclusion and discussion}
In conclusion, we have studied the three-state quantum Potts model by Monte Carlo simulations in the frame of SSE; the cluster algorithm we used is a direct generalization 
of that for the quantum Ising model.
The study includes the one and two dimensional ferromagnetic cases and the two dimensional antiferromagnetic case. 
The phase transition and critical phenomena are the same as the $d+1$ dimensional classical Potts model
in the sense of universality.

The classical antiferromagnetic Potts model has very interesting phase transitions and critical phenomena\cite{3slayer,4sAFPMC,mixedIsing,entropy-driven1,entropy-driven2},
especially in two dimensions. Generally, the symmetry of the
order parameter is determined not only by the symmetry of the Potts spin but also the lattice structure, for which the $Z_6$ and `emergent' O(2) symmetry is a good example;
therefore the model may have different phases and phase transitions on different lattices for the same $q$. It is an interesting question if the quantum fluctuation is 
introduced for the antiferromagnetic Potts model.
For example, in Ref. \onlinecite{mixedPotts} the mixed Potts model with $q=4$, 5, and 6 are also studied, where an `emergent' O(n) symmetry is found, with $n=q-1$. 
It is very possible that it is also the case for the antiferromagnetic quantum Potts model on the square lattice; we leave this question for further study.

\section{Acknowledgement}
Ding is supported by the Anhui Provincial Natural Science Foundation under Grant Numbers 1508085QA05 and 1408085MA19.
Deng is supported by the National Science Foundation of China (NSFC) under Grant Number 11625522.

\end{document}